\documentclass[aps,prl,preprint,groupedaddress]{revtex4}

\usepackage{graphicx}
\usepackage{dcolumn}
\usepackage{bm}

\begin{document}


\title{The evaluation of protein folding rate constant is improved by predicting	
       the folding kinetic order with a SVM-based method}


\author{Emidio Capriotti, Rita Casadio}
\affiliation{\small Biocomputing Group, Dept of Biology/CIRB University of Bologna, ITALY\\
	e-mail: emidio@biocomp.unibo.it
	}



\begin{abstract}
Protein folding is a problem of large interest 
since it concerns the mechanism by which the genetic information 
is translated into proteins with well defined three-dimensional 
(3D) structures and functions. Recent data on protein folding 
suggest that several pathologies such as prion and Alzheimer 
diseases may be due to change in protein stability during folding 
processes. Recently theoretical models have been developed to 
predict the protein folding rate considering the relationships 
of the process with tolopological parameters derived from the 
native (atomic-solved) protein structures. Previous works classified 
proteins in two different groups exhibiting either a single-exponential 
or a multi-exponential folding kinetics. It is well known that 
these two classes of proteins are related to different protein 
structural features. The increasing number of available experimental 
kinetic data allows the application to the problem of a machine 
learning approach, in order to predict the kinetic order of the 
folding process starting from the experimental data so far collected. 
This information can be used to improve the prediction of the 
folding rate.

In this work first we describe a support vector machine-based 
method (SVM-KO) to predict for a given protein the kinetic order 
of the folding process. Using this method we can classify correctly 
78\% of the folding mechanisms over a set of 63 experimental 
data. Secondly we focus on the prediction of the logarithm of 
the folding rate. This value can be obtained as a linear regression 
task with a SVM-based method. In this paper we show that linear 
correlation of the predicted with experimental data can improve 
when the regression task is computed over two different sets, 
instead of one, each of them composed by the proteins with a 
correctly predicted two state or multistate kinetic order.\\
\textbf{Keywods:} folding kinetics, kinetic order, folding rate, 
machine learning, support vector machine, contact order.
\end{abstract}


\maketitle

\section{1 Introduction}
In the last years, many theoretical and experimental studies 
have focused on the problem of describing the mechanism of protein 
folding \cite{jack,Plax,Fer,Gianni,compa,Plax00,Gar}. An important result was the development of empirical 
models that estimate protein folding kinetics and rates. The 
number of proteins under investigation is rapidly increasing, 
allowing more data to be collected. Many proteins fold by a simple 
two-state transition mechanism (TS), lacking observable folding 
intermediates under any experimental condition. In turn, other 
proteins are endowed with intermediates during the folding process; 
their folding process is therefore classified as a multistate 
one (MS). 

Experimental and theoretical work focused particularly on small 
two-state folding (TS) proteins. It was demonstrated that the 
logarithm of the in-water folding rates of these proteins correlates 
with some topological parameter as computed from their 3D structure 
or from that of closely related proteins, such as single point 
mutants or homologs with high level of sequence identity \cite{jack,zz,ivan}. 
Other methods predict protein folding rates starting from the 
Einstein diffusion equation \cite{debe} or from the secondary structure 
of the protein \cite{gong}. More recent work demonstrated that the chain 
length is one of the main determinants of the folding rate for 
proteins with a multistate folding (MS) kinetics \cite{gal,ivanas}. As 
a general observation, it appears that the logarithm of the folding 
rate correlates with structural topological parameters in TS 
proteins and with chain length in MS proteins. \\
In this paper we adopt a different perspective: we use the experimental 
data so far collected and, based on these observations, we develop 
a method to predict salient aspects of protein folding that can 
be directly computed starting from the protein structure.\\

\section{2 Material and Methods}
The problem here addressed concerns the kinetics and mechanism 
of the protein folding: starting from few simple parameters derived 
from the protein structure, the aim is to predict important features 
of the folding mechanism. In particular we implement a support 
vector machine (SVM)-based method trained over a set of 63 proteins 
known with atomic resolution and whose folding pathway has been 
experimentally characterized to predict the logarithm of the 
folding rate and whether the protein folds through intermediate 
states or not.

\subsection{2.1 Database and Tools}
Our data set is derived from the supplementary material of \cite{ivanas}. 
It contains folding data determined for 63 proteins, 38 of which 
are endowed with a TS folding mechanism. The other 25 proteins 
have a MS folding mechanism. The set comprises only single-domain 
proteins having no S-S bonds and/or no covalently bound ligands. 
When necessary, we generate two different sets (TS1 and MS1) 
including only those proteins correctly predicted with the corresponding 
folding mechanisms (TS1 comprises 34 proteins and MS1 15 proteins). 

The in-water folding rates (k$_{f}$) and native structure of these 
proteins have been established experimentally. The protein structure 
were collected from the Protein Data Bank (www.pdb.org) \cite{pdb}.\\
The method proposed here predicts some features of the protein 
folding process using a SVM approach. In particular we choose 
the LIBSVM tools available online at the web site http://www.csie.ntu.edu.tw/\ensuremath{\sim}cjlin/libsvmtools/. 
\\
The protein secondary structure was calculated with the DSSP 
program (http://www.cmbi.kun.nl/gv/dssp/ \cite{dssp}).\\
Sequence clustering was performed by means of the \textit{blastclust} 
program available within the BLAST suite at http://www.ncbi.nlm.nih.gov/ 
\cite{blast}.
 
\subsection{2.2 Protein structural parameters }
In order to investigate the relationships between the folding 
rate constant and the protein native conformation we evaluate 
four structure-based parameters. The first parameter is the effective 
length of the protein chain (L$_{eff}$) defined as\\
\begin{equation}
 L_{eff} =L-L_{H} +3*N_{H} 
\end{equation}\\
where L is the chain length, L$_{H}$ is the number of residues in 
helical conformation and N$_{H}$ is the number of helices. The others 
topological parameters are: the contact order (CO), \\
\begin{equation}
CO=\frac{1}{N_{} N_{c} } \sum\limits_{k=1}^{N_{c} }\Delta L_{ij}  
\end{equation}\\
the absolute contact order\\
\begin{equation}
CO=\frac{1}{N_{c} } \sum\limits_{k=1}^{N_{c} }\Delta L_{ij}
\end{equation}\\
and the total contact distance\\
\begin{equation}
TCD=\frac{1}{N_{}^{2} } \sum\limits_{k=1}^{N_{c} }\Delta L_{ij}
\end{equation}\\
where N is the number of amino acid residues of a protein, N$_{c}$ 
is defined as total number of contacts and \ensuremath{\Delta}L$_{ij}$ = {\textbar}i-j{\textbar}. 

The number of contacts is evaluated considering all the residues 
that have two heavy atoms within a given value of cut-off radius 
R and at a given sequence separation (w).

\subsection{2.3 The predictor}
The method addresses two different tasks: (1) the prediction 
of the existence of intermediate states in protein folding and 
(2) the prediction of the logarithm of the folding rate value. 
The former case is a classification task, discriminating whether 
for a given protein the folding pathway is or is not endowed 
with intermediate states; the latter one in turn is a fitting-regression 
task for estimating the folding rate. To address the two tasks, 
we developed a method based on support vector machines and relying 
on the same input for testing different kernel functions. Also, 
different SVMs explore different protein features. SVMs take 
two inputs for a given protein: the chain length and, one at 
a time, the four structured-based parameters described above 
(Eqn 1-4). We found that the best performing predictor was the 
one having as input the protein chain length and the contact 
order, tested by splitting the dataset in five parts and adopting 
a cross-validation procedure. The methods were then optimized 
trying different values of cut-off radius (R) and of sequence 
separation (w). 

In the final section of our work we focus our attention on the 
49 proteins that are correctly classified as TS and MS proteins. 
According to the protein kinetic order we split the dataset in 
two subsets, composed by 34 TS proteins and 15 MS proteins, respectively. 
We show that when linear regression fitting is performed on the 
two sets independently, the prediction of the logarithm of the 
folding rate is largely improved (see below).

\subsection{2.4 Scoring the classification performance}
All the results obtained with our systems are scored using a 
cross-validation procedure on the data pertaining to the 63 proteins. 
The dataset was divided in 5 subsets, putting in the same set 
proteins with the same PDB code and proteins with related sequences, 
as obtained by means of the \textit{blastclust} program, by adopting 
the default value of length coverage equal to 0.9 and the score 
coverage threshold equal to 1.75.

The efficiency of the predictor is scored using the statistical 
indexes defined in the following. The overall accuracy is:\\
\begin{equation}
Q2=\frac{p}{N}
\end{equation}\\
where p is the total number of correctly predicted folding mechanisms 
and N is the total number of proteins. \\
The Matthews correlation coefficient MC is defined as: \\
\begin{equation}
MC(s)=\frac{p(s)n(s)-u(s)o(s) }{D}
\end{equation}\\
where D is the normalization factor [[p(s)+u(s)] [p(s)+o(s)] 
[n(s)+u(s)] [n(s)+o(s)]]$^{1/2}$, for each class s (TS and MS, for 
two-state and multistate folding processes, respectively); p(s) 
and n(s) are the total number of correct predictions and correctly 
rejected assignments, respectively, and u(s) and o(s) are the 
numbers of under and over predictions.\\
Finally, it is very important to assign a reliability score to 
each SVM prediction. Using one SVM output this is obtained by 
computing:\\
\begin{equation}
Rel(i)=20*abs\left[ O(i)-0.5\right]
\end{equation}

\subsection{2.5 Scoring the regression performance}
The quality of the prediction when evaluating the protein folding 
constant rates was assessed by computing the Pearson linear correlation 
coefficient r and the associated value of the standard error \ensuremath{\sigma}.\\

\section{3 Results and Discussion}
In order to solve the tasks discussed in section 2.3 we developed 
different support vector machines. Taking advantage of previous 
studies, each of the SVMs considers two important protein features: 
(1) sequence length and (2) the four structural parameters described 
above. The best performing predictor was then optimized testing 
different values of cut-off radius (R), different sequence separation 
values (w) and different kernel functions. We found that the 
best performance was achieved by a SVM endowed with a linear 
kernel function K(x$_{i}$,x$_{j}$)=x$_{i}$Tx$_{j}$ (data not shown).

\subsection{3.1 Structural parameter optimization}
Previous studies have highlighted in proteins the correlation 
between folding kinetics and structural parameters as described 
in section 2.2 \cite{Plax00,zz,ivan,ivanas}. Table I lists the scoring performance 
of each method when predicting the logarithm of the folding rate 
and the folding kinetics.

\begin{table}
\begin{ruledtabular}
\begin{tabular}{cccccc}

{ \textbf{{\small Structural Parameter}}} & 
{  } & 
{  \textbf{{\small L}}$_{\mathbf{eff}}$} & 
{  \textbf{{\small CO}}} & 
{  \textbf{{\small A}}\textbf{{\small CO}}} & 
{  \textbf{{\small TCD}}}\\
\hline
{\centering \textbf{{\small Prediction of }}} & 
{  \textbf{{\small MC}}} & 
{  {\small 0.15}} & 
{  {\small 0.42}} & 
{  {\small 0.27}} & 
{  {\small 0.36}}\\
\cline{2-6}
{\centering \textbf{{\small Folding States}}} & 
{  \textbf{{\small Q2}}} & 
{  {\small 57.1}} & 
{  {\small 73.2}} & 
{  {\small 65.9}} & 
{  {\small 69.8}}\\
\hline
{\textbf{{\small  {\centering Prediction of }}}} & 
{  \textbf{{\small r}}} & 
{  {\small 0.45}} & 
{  {\small 0.64}} & 
{  {\small 0.45}} & 
{  {\small 0.63}}\\
\cline{2-6}
{\textbf{{\small  \centering  log(k}$_{\mathbf{f}}$\textbf\small)}} & 
{  \textbf{{\small \ensuremath{\sigma}}}} & 
{  {\small 1.57}} & 
{  {\small 1.39}} & 
{  {\small 1.57}} & 
{  {\small 1.37}}\\
\end{tabular}
\end{ruledtabular}
\caption{\textbf{{\small Scoring the SVM method.}} {\small The first 
two rows list the accuracy (Q2) and the Matthew's correlation 
coefficient (MC) of the four methods that include in the SVM 
input one of the different structural parameters and the sequence 
length. The four SVMs labeled with the name of the relative structural 
parameter, are tested in the binary classification between two-state 
(TS) and multistate (MS) folding mechanism. In the last two rows 
the correlation coefficient (r) and the standard error (}{\small \ensuremath{\sigma}}{\small ) 
of the previous methods for the prediction of the logarithm of 
the folding rate (k}$_{f}${\small ) are reported.}}{}
\end{table}

\subsection{3.2 Optimization of the cut-off radius}

The results shown in Table I indicate that the best SVM method 
has as input the sequence length and the contact order (see column 
CO). For this method we tested different values of the cut-off 
radius. In Table II (see next page), the scoring indexes for the 
two previous tasks are shown as a function of the radius value 
ranging from 4 to 12 {\AA}.


\begin{table}
\begin{ruledtabular}
\begin{tabular}{cccccc}

{ \textbf{{\small Cut-off radius{\nobreakspace}({\AA})}}} & 
{  } & 
{  \textbf{{\textbf 4}}} & 
{  \textbf{{\textbf 6}}} & 
{  \textbf{{\textbf 9}}} & 
{  \textbf{{\textbf 12}}}\\
\hline
{ \textbf{{\small Prediction of }}} & 
{  \textbf{{\small MC}}} & 
{  {\small 0.42}} & 
{  {\small 0.31}} & 
{  {\small 0.48}} & 
{  {\small 0.40}}\\
\cline{2-6}
{ \textbf{{Folding States}}} & 
{  \textbf{{\small Q2}}} & 
{  {\small 73.2}} & 
{  {\small 67.1}} & 
{  {\small 75.6}} & 
{  {\small 72.0}}\\

\hline
{\textbf{{\small {\centering Prediction of }}}} & 
{  \textbf{{\small r}}} & 
{  {\small 0.64}} & 
{  {\small 0.61}} & 
{  {\small 0.65}} & 
{  {\small 0.64}}\\
\cline{2-6}
{\textbf{\small \centering  log(k}$_{\mathbf{f}}$\textbf\small )} & 
{  \textbf{{\ensuremath{\sigma}}}} & 
{  {\small 1.39}} & 
{  {\small 1.44}} & 
{  {\small 1.35}} & 
{  {\small 1.38}}\\
\end{tabular}
\end{ruledtabular}
\caption{ \textbf{{\small Scoring SVMs as a function of the cut-off 
radius.}} {\small Here the method takes as input protein sequence length 
and contact order. The last structural parameter is calculated} {\small using 
a cut-off radius ranging} {\small from 4 to 12 {\AA}. The first two 
rows list the quality of the prediction in the classification 
task; the last two rows show the quality of the prediction of 
the logarithm of the folding rate (k}$_{f}${\small ).}}{}
\end{table}

\subsection{3.3 Sequence separation optimization} 
When considering the protein folding mechanism, an important 
issue is the different contribution of local and non-local interactions. 
It is well known that local interactions involved in the formation 
of particular motifs of secondary structure are established between 
residues with a sequence separation below 4 residues that is 
approximately the span length of one turn of an \'{a}-helix structure. 
Therefore, by increasing the value of w in the calculation of 
CO, we should encompass local interactions and include also contacts 
between residues that may contribute to non-local interactions 
during the folding process. We address this task by evaluating 
the contact order as a function of sequence separation; the best 
performing implementation of SVMs was consequently optimized 
and the results are shown in Table III.

\begin{table}
\begin{ruledtabular}
\begin{tabular}{ccccccc}
{\textbf{\centering {\small Windows Length}}} & 
{  } & 
{\ {\small 0}} & 
{ {\small 2}} & 
{ {\small 4}} & 
{ {\small 6}} & 
{ {\small 8}}\\
\hline
{\textbf{\centering {\small Prediction of}}} & 
{  {\textbf MC}} & 
{  {\small 0.48}} & 
{  {\small 0.50}} & 
{  {\small 0.46}} & 
{  {\small 0.53}} & 
{  {\small 0.42}}\\
\cline{2-7}
{\textbf{\centering {\small  Folding States}}} & 
{  {\textbf Q2}} & 
{  {\small 75.6}} & 
{  {\small 76.1}} & 
{  {\small 74.6}} & 
{  {\small 77.7}} & 
{  {\small 73.1}}\\
\hline
{\textbf{{\small {{\centering Prediction of }}}}} & 
{  \textbf{{\textbf r}}} & 
{  {\small 0.65}} & 
{  {\small 0.60}} & 
{  {\small 0.61}} & 
{  {\small 0.58}} & 
{  {\small 0.60}}\\
\cline{2-7}
{\textbf{\small {\centering  log(k}$_{\mathbf{f}}$\textbf\small )}} & 
{  \textbf{{ \ensuremath{\sigma}}}} & 
{  {\small 1.35}} & 
{  {\small 1.52}} & 
{  {\small 1.29}} & 
{  {\small 1.45}} & 
{  {\small 1.41}}\\
\end{tabular}
\end{ruledtabular}
\caption{ \textbf{{\small Local vs global interactions.}} {\small In this 
table we report the accuracy of the best methods (cut-off radii 
9 {\AA}) for different values of a sequence separation (w), spanning 
from 0 to 8 residues, when evaluating the contact order number. 
In other words, we consider only contact between residue i and 
j provided that {\textbar}i-j{\textbar}\texttt{>}w. The efficiency of the 
predictions for the two tasks are scored using the same measures 
reported in Table I.}}{}
\end{table}

\subsection{3.4 Prediction of the folding mechanism.}
From our results we conclude that the best method for the binary 
classification between the two-state and the multistate folding 
mechanism takes as input the sequence length and the contact 
order. The best discrimination between TS and MS proteins is 
obtained when the contact order value is calculated considering 
a cut-off radius of 9 {\AA} and a sequence separation \texttt{>}6 residues. 
In figure 1 we report the accuracy (Q2) and the Matthew's correlation 
coefficient (MC) as a function of the reliability index (RI).

\begin{figure}
\includegraphics{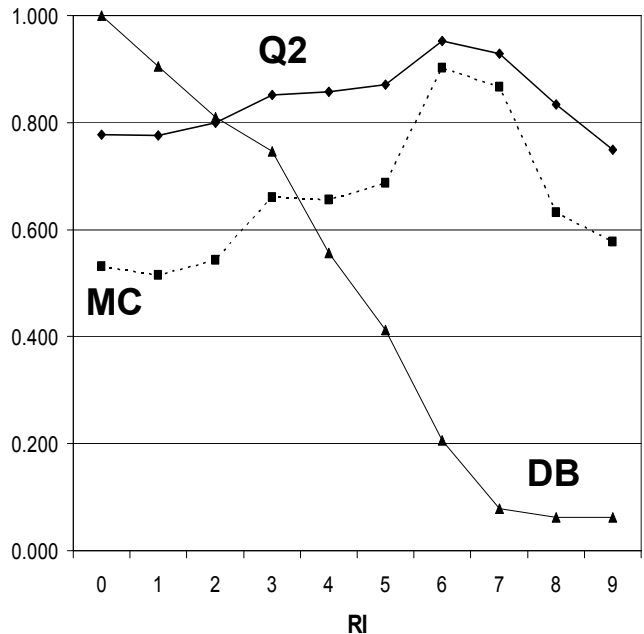}%
\caption{\small Accuracy (Q2) and Matthew's correlation 
coefficient (MC) as o function of the reliability index (RI). 
\small DB is the fraction of the dataset with a reliability value 
higher or equal to a given threshold.}
\end{figure}

\subsection{3.5 Prediction of the logarithm of the folding rate}
Similar to the classification task, the regression task for the 
prediction of the logarithm of the folding rate is optimized 
considering as input the sequence length and the contact order. 
The results of our method are shown in figure 2.

\begin{figure}
\includegraphics{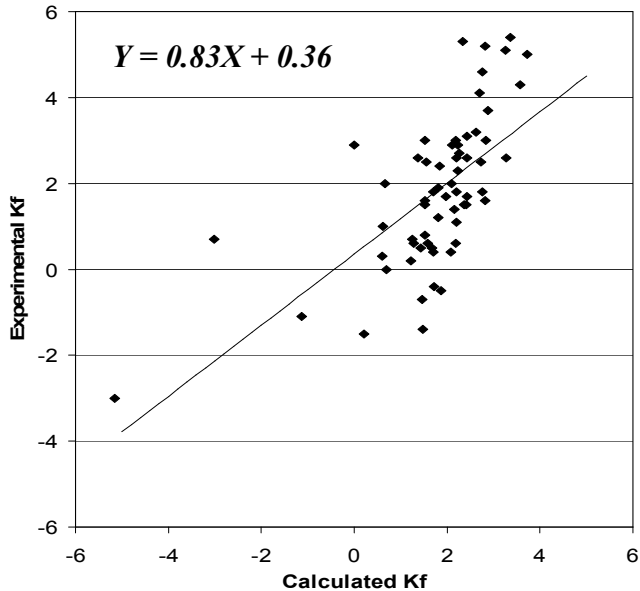}%
\caption{\small  Value of the logarithm of the folding rate 
(k$_{\mathbf{f}}$ \small ) versus its experimental value. The correlation 
coefficient for the best method previously described is 0.65 
and the standard error is 1.35. We also reported the equation 
of the linear best fit.}
\end{figure}

\subsection{3.6 Improving the prediction of folding rate}
Can we discriminate between two-state and multistate proteins 
when predicting the folding rate constant? In other words can 
we relate the folding rate constant to the protein structural 
features? To address this problem we selected from the dataset 
only those proteins for which our predictor correctly evaluates 
the kinetic order (49 proteins over 63 in the whole set) and 
we compute the linear regression between calculated and experimental 
folding rate for each of the two folding types. In figure 3 we 
report the regression plot calculated over the 34 that are correctly 
predicted as TS proteins.\\

\begin{figure}
\includegraphics{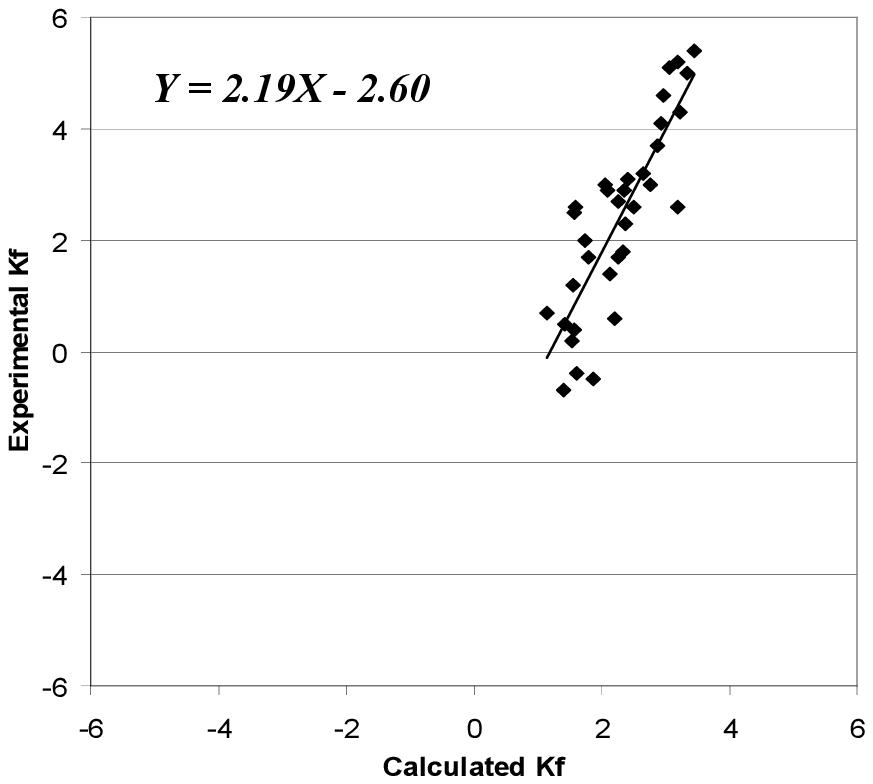}%
\caption{\small Value of the logarithm of the folding 
rate (k$_{\mathbf{f}}$ \small ) versus its experimental value for correctly predicted 
TS proteins. The correlation coefficient is 0.84 and the standard 
error is 0.90. We also reported the equation of the linear best 
fit. These values are calculated on the subset of 34 over 39 
correctly predicted TS proteins.}
\end{figure}

In the figure 4 the last procedure was performed for the 15 correctly 
predicted MS proteins
It appears that both the slope and the intercept are different, 
depending on the protein folding type, and that the correlation 
coefficient and standard error are better when each folding type 
is considered separately than when accumulated. \\

\begin{figure}
\includegraphics{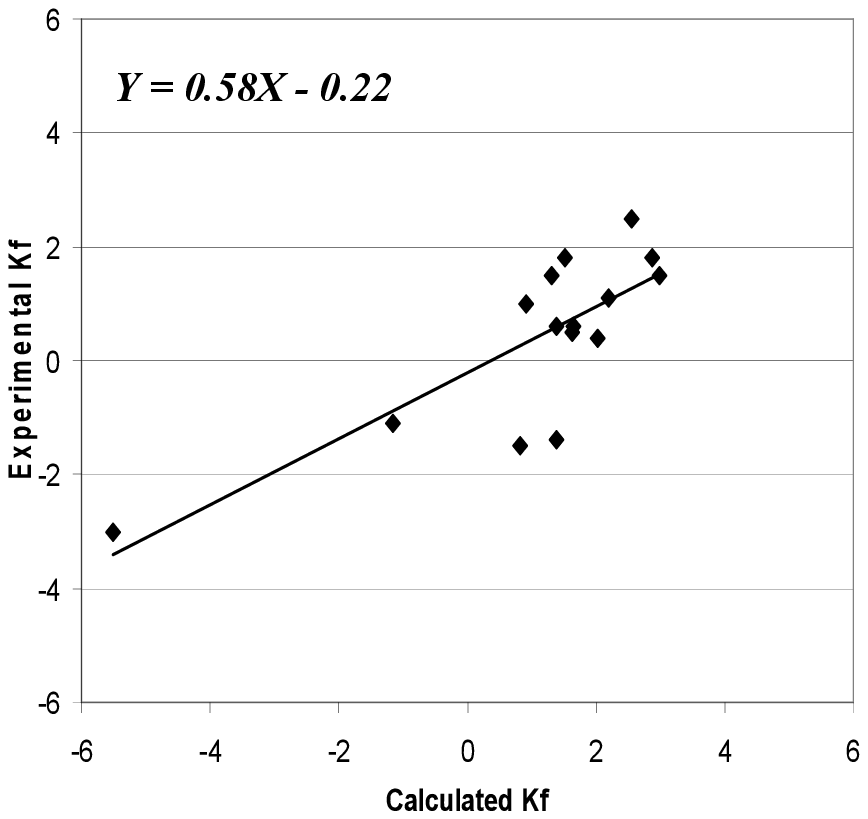}%
\caption{\small Value of the logarithm of the folding rate 
(k$_{\mathbf{f}}$ \small ) versus its experimental value for correctly 
predicted TS proteins. The correlation coefficient is 0.79 
and the standard error is 0.89. We also reported the equation 
of the linear best fit. These values are calculated on the subset 
of 15 over 25 correctly predicted MS proteins.}
\end{figure}

\section{4 Conclusion}
This work represents a first attempt to address the problem of 
the prediction of the folding mechanism using a machine learning 
approach. In particular we try to predict whether the folding 
process follows a two-state or a multistate mechanism and the 
logarithm of the folding rate considering only few simple inputs: 
the length of the protein sequence and the contact order, as 
calculated according to the eq. (2). This is the first time, 
at the best of our knowledge, that a statistical evaluation of 
the problem is provided. We optimize our method considering different 
values of the cut-off radius and introducing a sequence separation 
for the calculation of the contact order (CO) from the protein 
structure, in order to discriminate local versus non local interactions. 
Our approach allows to generalize on the given examples since 
it is tested adopting a cross-validation procedure. We find that 
the best predictive performance is achieved when the value of 
the contact order is calculated using a cut-off radii of 9 {\AA} 
and a sequence separation larger or equal to 6, suggesting that 
non local more than local interactions are important in determining 
the parameters at hand for the given protein set. \\
With our method the prediction of possible intermediate states 
during the folding process reaches accuracy of 78\% with a significant 
Matthew's correlation coefficient of 0.53. Furthermore, when 
predictions with a reliability index value =3 are considered, 
the SVM method increases its accuracy to 85\% and its correlation 
to 0.66 over 75\% of the database. Results in Tab. 2 indicate 
that for discriminating between TS and MS folding mechanisms, 
contacts between residues with sequence separation \texttt{>}6 are 
important. In turn, for predicting the value of the logarithm 
of the folding rate the highest score is obtained considering 
all the contacts. On the contrary, with respect to the classification 
between TS and MS proteins, the regression task for the prediction 
of the logarithm of k$_{f}$, performs better when local and non 
local interactions are considered taking also into account contacts 
with sequence separation less or equal then 6. In this particular 
task our best method reaches a significant correlation coefficient 
of 0.65 with a related standard error of 1.35. These values can 
be considered satisfactory, since they are obtained with only 
two element vectors as input in the training of the SVM and since 
the method is tested using a cross-validation procedure.\\
In order to improve the prediction of the logarithm of the folding 
rate in the last part of this work we developed a new method 
that use the two different linear regression calculated over 
two subsets. The two sets are built starting from the 63 proteins 
and selecting the 34 proteins with correctly predicted TS kinetic 
and the 15 proteins with correctly predicted MS kinetic. On this 
set of 49 proteins data the linear regression fit calculated 
between experimental and predicted logarithm of k$_{f}$ is 0.74 
and the standard error is 1.25. The subsequent division of the 
database in predicted TS and MS proteins leads to an increase 
of the correlation between experimental and predicted logarithm 
of k$_{f}$. The obtained values are 0.84 when the regression fitting 
is performed on the 34 correctly predicted TS proteins and 0.79 
when the regression fitting is calculated over the 15 predicted 
MS proteins, respectively. Similarly, a general decrease of the 
standard error that reaches a mean value of about 0.90 is noticed. 
\\
In spite of the fact that the SVM predicts incorrectly 22\% of 
the folding kinetic order values, the new method allows us to 
improves the efficacy in the evaluation of the folding rate. 
The results confirm the hypothesis that the folding rate of TS 
and MS are a function of the same protein structural features 
(protein length and contact order). However the relative weight 
of their contribution to the overall rate constant may be different 
for the two folding types.\\
This work represents to our knowledge the first attempt to predict 
the folding protein type from sequence length and contact order, 
as computed from the protein structure. Furthermore we can also 
predict the folding rate constant. This value is better correlated 
to experimental values when the two folding protein types are 
considered separately, suggesting that sequence length and contact 
order interplay differently as a function of the protein folding 
type. Although the number of available experiments is not so 
high this research suggests some ideas and a general procedure 
to investigate the kinetic of the protein folding.

\section{ACKNOWLEDGMENTS}
This work was supported by following grants: PNR 2001-2003 (FIRB art. 8) and PNR 2003 project (FIRB art. 8) on Bioinformatics for Genomics and Proteomics and LIBI-Laboratorio Internazionale di Bioinformatica, both delivered to RC. EC is supported by a grant of the European Union VI Framework Programme to the Bologna Node of the Biosapiens Network of Excellence project.

\newpage
\bibliography{prova}
\end{document}